\documentclass[fleqn,usenatbib]{mnras}
\usepackage{newtxtext,newtxmath}
\usepackage[T1]{fontenc}
\usepackage{ae,aecompl}
\usepackage{graphicx}	
\usepackage{amsmath}	
\usepackage{amssymb}	

\newcommand{\Ho}{67.8\pm1.3}

\title[First Cosmological Results using SNe Ia from DES: Measurement of $H_0$]{First Cosmological Results using Type Ia Supernovae from the Dark Energy Survey: Measurement of the Hubble constant}

\author[E.~Macaulay et al.]{
\parbox{\textwidth}{
\Large
E.~Macaulay$^{1}$ \thanks{email: \href{mailto:e.macaulay@port.ac.uk}{\nolinkurl{edward.macaulay@port.ac.uk}}},
R.~C.~Nichol,$^{1}$
D.~Bacon,$^{1}$
D.~Brout,$^{2}$
T.~M.~Davis,$^{3}$
B.~Zhang,$^{4,5}$
B.~A.~Bassett,$^{6,7}$
D.~Scolnic,$^{8}$
A.~M\"oller,$^{4,5}$
C.~B.~D'Andrea,$^{2}$
S.~R.~Hinton,$^{3}$
% ...
\begin{center} (DES Collaboration) \end{center}
}
\vspace{0.4cm}
\\
\parbox{\textwidth}{
Author affiliations are shown in Appendix \ref{appendix:affiliations}\\
}
}

\date{Accepted 2019 April 03. Received 2019 April 02; in original form 2018 November 08}

\pubyear{2019}

\begin{document}
\label{firstpage}
\pagerange{\pageref{firstpage}--\pageref{lastpage}}
\maketitle

\begin{abstract}
We present an improved measurement of the Hubble constant ($H_0$) using the `inverse distance ladder' method, which adds the information from \textcolor{black}{207} Type Ia supernovae (SNe Ia) from the Dark Energy Survey (DES) at redshift $0.018<z<0.85$ to existing distance measurements of \textcolor{black}{122} low redshift ($z< 0.07 $) SNe Ia (Low-$z$) and measurements of Baryon Acoustic Oscillations (BAOs).  Whereas traditional measurements of $H_0$ with SNe Ia use a distance ladder of parallax and Cepheid variable stars, the inverse distance ladder relies on absolute distance measurements from the BAOs to calibrate the intrinsic magnitude of the SNe Ia.  We find $H_0= \Ho $ km s$^{-1}$ Mpc$^{-1}$ (statistical and systematic uncertainties, 68\% confidence).  Our measurement makes minimal assumptions about the underlying cosmological model, and our analysis was blinded to reduce confirmation bias.  We examine possible systematic uncertainties and all are below the statistical uncertainties.  Our $H_0$ value is consistent with estimates derived from the Cosmic Microwave Background assuming a $\Lambda$CDM universe \citep{2018arXiv180706209P}. 
\end{abstract}


\section{Introduction}
\label{sec:intro}

The precise value of the Hubble constant ($H_0$) has again become one of the most debated topics in cosmology \citep[see][]{freeman17}. This debate has been fuelled by the apparent disagreement between local, direct measurements of $H_0$, primarily \cite{2016ApJ...826...56R} who find $H_0=73.24\pm1.74$ km s$^{-1}$ Mpc$^{-1}$, and  estimates derived from the Cosmic Microwave Background (CMB) which give $H_0=67.4\pm0.5$ km s$^{-1}$ Mpc$^{-1}$ \citep{2018arXiv180706209P}, assuming a $\Lambda$CDM universe. This discrepancy has increased to $3.7\sigma$ with new parallax measurements to Cepheid variable stars by \cite{2018ApJ...855..136R} giving $H_0=73.48\pm1.66$ km s$^{-1}$ Mpc$^{-1}$.

This tension between the local measurements of the Hubble constant and the Planck+$\Lambda$CDM expectation may be due to unknown systematic uncertainties in the various observations, flaws in the theoretical assumptions, and/or under-estimation of the uncertainties on the measurements of $H_0$ \citep[e.g., see discussion in ][]{2017MNRAS.471.2254Z}.
  
Sample or cosmic variance has been proposed as a potential systematic effect for direct measurements of $H_0$. Cepheid variable stars can only be observed in the nearest galaxies, and the number of such galaxies that also have a well-determined SN Ia with which to calibrate the SN Ia luminosity zeropoint is small.  Thus these measurements only probe a small cosmological volume with a low number of galaxies for cross-calibration. However, \cite{2017arXiv170609723W} used $N$-body simulations to evaluate the sample variance, and found that it contributes a dispersion of 0.31 km s$^{-1}$ Mpc$^{-1}$ to the local measurements of $H_0$, which is too small to account for the discrepancy with Planck.

The discrepancy in $H_0$ may alternatively be due to physics beyond the $\Lambda$CDM model \citep{2016JCAP...10..019B,2016PhLB..761..242D,2016ApJ...826...56R,2017arXiv171002153D}.  A negative curvature ($\Omega_k$ \textless \, 0) could  account for the discrepancy, which would have implications for models of cosmic inflation \citep{2014PhRvD..89j3502D,2016MNRAS.463.1416G,2017ApJ...835...26F}.   Modifications to gravity could cause a larger acceleration than expected in $\Lambda$CDM \citep{2016PhRvD..94d3518P,2017PhRvD..96d3503D,2017NatAs...1..627Z}.  Alternatively, an additional relativistic species at the CMB epoch could account for the tension \citep{2012JCAP...07..053M,2017arXiv170108172V}.  We note however that explanations for the $H_0$ tension involving modified gravity, or an extra relativistic species, would increase tensions in measurements of $\sigma_8$ (the amplitude of the matter power spectrum) and expectations from Planck+$\Lambda$CDM \citep[e.g.,][]{2013PhRvL.111p1301M,2014PhRvD..89h3517Y,2017MNRAS.471.1259J,2017JCAP...01..028C}.

This tension has motivated the development of new, independent ways to measure $H_0$. For example, \cite{2018arXiv180901274B} find $H_0=72.5^{+2.1}_{-2.3}$ km s$^{-1}$ Mpc$^{-1}$ from measurements of time delays from strongly lensed quasars, and \cite{2017ApJ...851L..36G} find $H_0=75.5^{+11.6}_{-9.6}$ km s$^{-1}$ Mpc$^{-1}$ by using the gravitational wave event GW170817 as a standard siren.

In this paper, we present a new measurement of $H_0$ using spectroscopically-confirmed SNe Ia from the Dark Energy Survey \citep[see ][for details]{2015AJ....150..150F,2018arXiv180103181A}.  Since SNe Ia are relative, not absolute, distance indicators, their intrinsic magnitude must be calibrated using an absolute distance measurement.  This is the motivation behind the conventional distance ladder approach of calibrating local SNe Ia using Cepheid variable stars and parallax.  The approach we take is to calibrate the intrinsic magnitude of SNe Ia against the absolute distance measurements from the Baryon Acoustic Oscillations (BAOs) at $z>0.1$ (assuming the sound horizon from the CMB).  We then use the calibrated SN Ia distances to trace the expansion history of the Universe back to $z=0$ to determine $H_0$.  

We note that although BAO measurements alone could derive a value for $H_0$ (e.g., see Figure \ref{fig:inv_dist_ladder}), it would rely on the assumption of a cosmological model to extrapolate the BAO measurements to $z=0$ \citep[see e.g.,][]{rozo17}. By using calibrated SNe Ia across a range of redshifts, we can determine $H_0$ more directly without assuming a specific cosmological model such as $\Lambda$CDM.

$H_0$ was first measured using this `inverse distance ladder' technique by \cite{2015PhRvD..92l3516A}, who found $H_0=67.3\pm1.1$ km s$^{-1}$ Mpc$^{-1}$ with SNe Ia from the Joint Light Curve Analysis \citep[JLA;][]{2014A&A...568A..22B} and BAO measurements from the Baryon Oscillation Spectroscopic Survey (BOSS) Data Release Eleven (DR11).  This result was updated with BOSS DR12 in \cite{2017MNRAS.470.2617A}, finding $H_0=67.3\pm1.0$ km s$^{-1}$ Mpc$^{-1}$.  

In this paper, we use \textcolor{black}{207} new, spectroscopically-confirmed SNe Ia from the DES Supernova Program (DES-SN3YR) to measure $H_0$ with this inverse distance ladder technique. While this sample contains fewer supernovae than JLA, the DES-SN3YR sample has the key advantage of spanning the entire redshift range of the available galaxy BAO measurements (e.g., $z_{\rm eff}=0.112$ to $0.61$) in a single survey. This is not true of other inverse distance ladder measurements which rely on the JLA sample, because different SN surveys must be combined in order to cover this redshift range (e.g., the SDSS SN sample at $z\simeq0.1$, SNLS at $z\simeq0.6$). 

We describe the data and method used for our analysis in Section \ref{sec:data}, and our results in Section \ref{sec:results}. We conclude in Section \ref{sec:Discussion}.

\section{Methodology and Data}
\label{sec:data}

We use a similar methodology as \cite{2015PhRvD..92l3516A}, using BAO distance measurements to calibrate the SNe Ia.  This breaks the well-known degeneracy between the SNe Ia peak absolute magnitude and $H_0$. While BAO data alone can constrain $H_0$, these measurements typically assume a specific cosmological model \citep[e.g., a cosmological constant as in][]{rozo17}, since the BAO measurements do not have sufficient redshift coverage to determine $H(z)$ on their own (Figure \ref{fig:inv_dist_ladder} illustrates this point).  By combining BAO and SNe Ia, we can relax the assumption of a specific cosmological model when determining $H_0$. 

However, we do still require some model for the redshift-distance relationship to extrapolate these data to $z=0$. 
For this work, we adopt a cosmographical approach for the redshift-distance relationship, which is a smooth Taylor expansion about redshift, that makes minimal assumptions about the underlying cosmological model \citep{muth16,Zhang17,Feeney18}. We use Equations 6, 7 and 8 in \cite{Zhang17} to determine the luminosity distance $D_L(z)$ and Hubble parameter $H(z)$ as a function of redshift.  $D_L(z)$ is given by
\begin{equation}
\label{eq:luminosity}
D_L(z) = z + C_1 z^2 + C_2 z^3 + C_3 z^4 + C_4 z^5 + ... ,
\end{equation}
where
\begin{equation}
\label{eq:C1}
C_1 = \frac{1}{2}(1-q_0),
\end{equation}
\begin{equation}
\label{eq:C2}
C_2 = -\frac{1}{6}(1-q_0 - 3 q_0^2 + j_0),
\end{equation}
\begin{equation}
\label{eq:C3}
C_3 = \frac{1}{24}(2-2q_0 - 15 q_0^2 + 5 j_0 + 10 q_0 j_0 + s_0),
\end{equation}
and
\begin{equation}
\begin{split}
\label{eq:C4}
C_4 &= \frac{1}{120}(-6+6q_0 + 81 q_0^2 + 165 q_0^3 + 105q_0^4 +10 j_0^2 - 27j_0  \\
& - 110 q_0 j_0 - 105 q_0^2 j_0 - 15 q_0 s_0 -11 s_0 - l_0).
\end{split}
\end{equation}
$H(z)$ is given by
\begin{equation}
\begin{split}
\label{eq:Hz}
H(z) &=  H_0 [ 1 + (1+q_0)z + \frac{1}{2}(-q_0^2 + j_0) z^2 \\
     & +\frac{1}{6}  (3q_0^2 + 3q_0^3 - 4q_0j_0 - 3j_0 - s_0) z^3 \\
     & +\frac{1}{24} ( -12q_0^2 - 24 q_0^3 - 15q_0^4 + 32 q_0 j_0 + 25 q_0^2 j_0 \\
     & +7q_0 s_0 + 12j_0 -4j_0^2 + 8s_0 + l_0 ) z^4 ] + ... 
\end{split}
\end{equation}
We find that including the lerk ($l_0$) parameter ($z^5$) in our cosmographic model increases the Bayesian Information Criterion from 40.4 to 44.1, which indicates that including this additional parameter is not warranted by the data. The fitting parameters in this Taylor expansion are then $H_0$ (Hubble constant), $q_0$ (deceleration), $j_0$ (jerk), and $s_0$ (snap).  This is consistent with \cite{Gomez18} who found that fourth-order polynomials and above made minimal improvement to the Bayesian and Akaike information criteria when fitting a larger set of data including both SNe Ia and other cosmological data-sets. They also noted that such higher-order polynomials had minimal effect on the value of $H_0$ they determined. We assume uniform priors on these cosmographical parameters and therefore, our results are relatively insensitive to the details of the assumed underlying late-time redshift-distance relationship.

Throughout, we must assume the validity of the cosmic distance-duality relation; that the luminosity distance $D_L(z)$ is related to the angular diameter distance $D_A(z)$ by $D_L(z) = D_A(z)(1+z)^2$.  This  well-known relationship in cosmology is applicable to general metric theories of gravity in which photons are conserved and travel on null geodesics \citep[e.g., see][]{2004ApJ...607..661B}.

To determine $H_0$, we perform a combined analysis of SNe Ia and BAO with a Gaussian prior on $r_s$ (the sound horizon at recombination) based on CMB data. All these data are required and complementary, and assumed to be independent. The individual likelihood functions are assumed to be Gaussian and given by 
\begin{equation}
\label{eq:likelihood}
\ln \mathcal{L}_{x} = \Delta_{x}^T \mathsf{C}^{-1} \Delta_{x},
\end{equation}
\noindent where $x$ above is either SNe Ia or BAO data. $\mathsf{C}$ is the data covariance matrix for either data-set, and $\Delta_{x}$ is the difference vector between the data-sets and their corresponding values in the cosmographical model.

These likelihood functions are then combined to give
\begin{equation}
\label{eq:likelihood_combined}
\ln \mathcal{L}(\Theta,M^1_B,r_s) = \ln \mathcal{L}_{\rm{SN}}(\Theta,M^1_B) +\ln \mathcal{L}_{\rm{\rm{BAO}}}(\Theta,r_s) + \ln \mathcal{L}_{r_s}(r_s),
\end{equation}
where $\Theta= \left[ H_0,q_0,j_0,s_0   \right]$ is a common set of cosmographic parameters and $M^1_B$ is the SNe Ia absolute magnitude at peak (see Section \ref{sec:sn}).

\subsection{Baryon Acoustic Oscillations}
\label{sec:BAO}

The scale of the BAO is a well-established cosmological standard ruler \citep[e.g.,][]{2003ApJ...594..665B,2003ApJ...598..720S,2005ApJ...633..560E,2011MNRAS.418.1707B,2013A&A...552A..96B,2014MNRAS.441...24A,2017MNRAS.470.2617A}.  With the physical scale set by the sound horizon at recombination ($r_s$), BAOs provide absolute distance measurements over a range of redshifts.  In order to measure the BAO signal from galaxy redshift surveys, a fiducial cosmology must be assumed in order to convert the observed angles and redshifts into distances. 

We emphasise that this does not imply that a BAO measurement is limited to a consistency test of that assumed fiducial cosmology, since most BAO analyses typically fit for $\alpha^{\rm{BAO}}$, a dimensionless parameter measuring the ratio of the observed BAO scale to the scale expected in the fiducial cosmology.

An isotropic BAO analysis, where the BAO signal is measured from pairs of galaxies averaged over all angles, is sensitive to the volume averaged distance \citep[e.g., $D_V$, see ][]{2015PhRvD..92l3516A}.  We can relate the expected $D_V(\Theta,z)$ to the observed $\alpha^{\rm{BAO}}$ and the fiducial BAO values by
\begin{equation}
\label{eq:Dv}
\frac{D_V(\Theta,z)}{r_s}=\alpha^{\rm{BAO}} \frac{D^{\rm{fid}}_V(z)}{r^{\rm{fid}}_s},
\end{equation}
where
\begin{equation}
\label{eq:Dv}
D_V(\Theta,z)=[z D_H(\Theta,z)D^2_M(\Theta,z)]^{1/3}.
\end{equation}
$D_H(\Theta,z)$ is the Hubble distance, given by
\begin{equation}
\label{eq:Dv}
D_H(\Theta,z) = \frac{c}{H(\Theta,z)}, 
\end{equation}
and $D_M(\Theta,z)$ is the comoving angular diameter distance. 

In our likelihood analysis, we use the observed 
$D_{V}(z_{\mathrm{eff}}=0.122)=539\pm 17(r_{s}/r^{\mathrm{fid}}_{s})\,\mathrm{Mpc}$ (68\% confidence) taken from \cite{carter18}, based on a re-analysis of the 6-degree Field Galaxy Survey \citep{2011MNRAS.416.3017B} and Sloan Digital Sky Survey Main Galaxy Sample \citep{2015MNRAS.449..835R}.  

At higher redshift, we use the `Consensus' BOSS DR12 data-set from \cite{2017MNRAS.470.2617A} which provides a two-dimensional description of the clustering, dividing the separation between pairs of galaxies into components across and along the line-of-sight, now summarising the clustering with two parameters of $\alpha_{\perp}^{\rm BAO}$ and $\alpha_{||}^{\rm BAO}$, respectively.  Their expected values can now be related to their observed values by
\begin{equation}
\label{eq:Dv}
\frac{D_M(\Theta,z)}{r_s}=\alpha_{\perp}^{\rm{BAO}} \frac{D^{\rm{fid}}_M(z)}{r^{\rm{fid}}_s},
\end{equation}
and
\begin{equation}
\label{eq:Dv}
\frac{D_H(\Theta,z)}{r_s}=\alpha_{||}^{\rm{BAO}} \frac{D^{\rm{fid}}_H(z)}{r^{\rm{fid}}_s}.
\end{equation}
The BOSS DR12 data-set consists of measurements of $D_M(z)$ and $H(z)$ at three effective redshifts of $z_{\rm{eff}}=[0.38,0.51,0.61]$ (6 measurements in total).  The covariance matrix for these six measurements includes the correlation between $D_M(z)$ and $H(z)$ at each $z_{\rm{eff}}$, and the correlation between these six measurements in different redshift bins. 

In Equation \ref{eq:likelihood}, $\ln \mathcal{L}_{\rm{\rm{BAO}}}(\Theta,r_s)$ is then the combined likelihood of the two BAO data-sets from \cite{carter18} and BOSS DR12. This likelihood requires knowledge of the sound horizon at recombination ($r_s$), which depends on the sound speed at these earlier epochs and thus the baryon density ($\omega_b$) and the total matter density ($\omega_{cb}$) in the early universe \citep[see Equation 16 of][]{2015PhRvD..92l3516A}. 

In our analysis, we adopt a Gaussian prior on $r_s$ of $147.05\pm0.30$ Mpc (68\% confidence) taken from the Planck 2018 analysis (\texttt{TT,TE,EE+lowE} result in Table 2).  By using the value of $r_s$ derived from only the \texttt{TT,TE,EE+lowE} Planck data, we minimise our sensitivity to physics of the late-time universe. Of these effects, CMB lensing is the most significant, although \cite{Feeney18} note that even including CMB lensing changes their value of $H_0$ by less than the statistical uncertainty on this measurement.

We explore the sensitivity of our results to this prior in Appendix \ref{sec:Discussion}, but note that the Planck measurement of $r_s$ comes from their measurement of the baryon and total matter densities, which in turn are only related to the heights of the acoustic peaks in the CMB power spectrum and not on their angular locations. Therefore, any dependencies introduced because of this Planck prior are based only on our correct understanding of plasma physics in the pre-recombination epoch, rather than assumptions about curvature and late-time dark energy, which are negligible in the early Universe for many cosmological models. 

\subsection{Type Ia Supernovae}
\label{sec:sn}
Type Ia supernovae are cosmological standard candles \citep[e.g.,][]{1998AJ....116.1009R,1999ApJ...517..565P,2009ApJ...700..331H,2009ApJS..185...32K,2010MNRAS.406..782S}. In this analysis, we use the DES-SN3YR sample of \textcolor{black}{207} new, spectroscopically-confirmed SNe Ia from the first three years of DES, which are  supplemented by \textcolor{black}{122} SNe Ia from the  CfA3 \citep{2009ApJ...700..331H}, CfA4 \citep{2012ApJS..200...12H} and CSP Low-$z$ sample  \citep{2006PASP..118....2H,2010AJ....139..519C,2010AJ....139..120F,2011AJ....142..156S,2017AJ....154..211K} ($z<0.09$) described in part in \cite{Scolnic2015}.

The details of the DES-SN3YR sample are provided in a series of papers as part of the overall DES-SN 3 year cosmology paper by \cite{2018arXiv181102374D}.  \cite{2018arXiv181109565D} describes the spectroscopic follow-up observations, \cite{2018arXiv181102378B} outlines the supernova scene model photometry, \cite{2018arXiv181102380L} details the DES photometric corrections, \cite{2019MNRAS.tmp..472K} presents the survey simulations and selection function.  \cite{2018arXiv181102377B} presents validations of the sample, systematic uncertainties, light-curve fits, and distance measurements. 

Our analysis uses the distances and covariance matrices derived in \cite{2018arXiv181102377B} with the `BBC' (BEAMS with Bias Correction) method \citep{2017ApJ...836...56K}. The distances have been binned in 18 redshift bins (originally 20 bins, but with 2 empty bins).

The distance modulus, $\mu$, for these supernovae is given by
\begin{equation}
\mu (\Theta,z) = 25 + 5 \log_{10}\left( D_{L} (\Theta,z)   \right),
\label{mu_theory}
\end{equation}
where $D_{L}(\Theta,z)$ is the luminosity distance. We relate $\mu$ to the observed SALT2 \citep{2007A&A...466...11G} light curve parameters by 
\begin{equation}
\mu = m_B- M^1_B + \alpha X_1 - \beta C+\Delta m_{\rm{host}}+\Delta B,
\label{mu_obs}
 \end{equation}
 where $m_B$ is the observed $B$-band peak magnitude, $M^1_B$ is the absolute magnitude of the SNe Ia, $X_1$ is the stretch parameter of the light curve, and $C$ is the colour parameter. $\alpha$ and $\beta$ are free parameters which are fitted for when calculating the distances.  Also, $\Delta m_{\rm{host}}$ is a correction applied for  host galaxy masses of $M_{\rm{stellar}}>10^{10} M_{\odot}$  \citep[see also][]{2010MNRAS.406..782S,2010ApJ...715..743K,2010ApJ...722..566L}. The stellar mass measurements for the host galaxies were obtained from fits to the DES SV galaxy photometry with the galaxy evolution modelling code ZPEG \citep{2002A&A...386..446L}  (see Smith et al. 2019, in prep. for details). $\Delta B$ is the expected $\mu$-correction due to the survey selection function for both the DES-SN3YR sample as discussed in detail in \cite{2018arXiv181102377B}.

\section{Results}
\label{sec:results}

\begin{figure*}
\begin{center}
 \includegraphics[width=18cm]{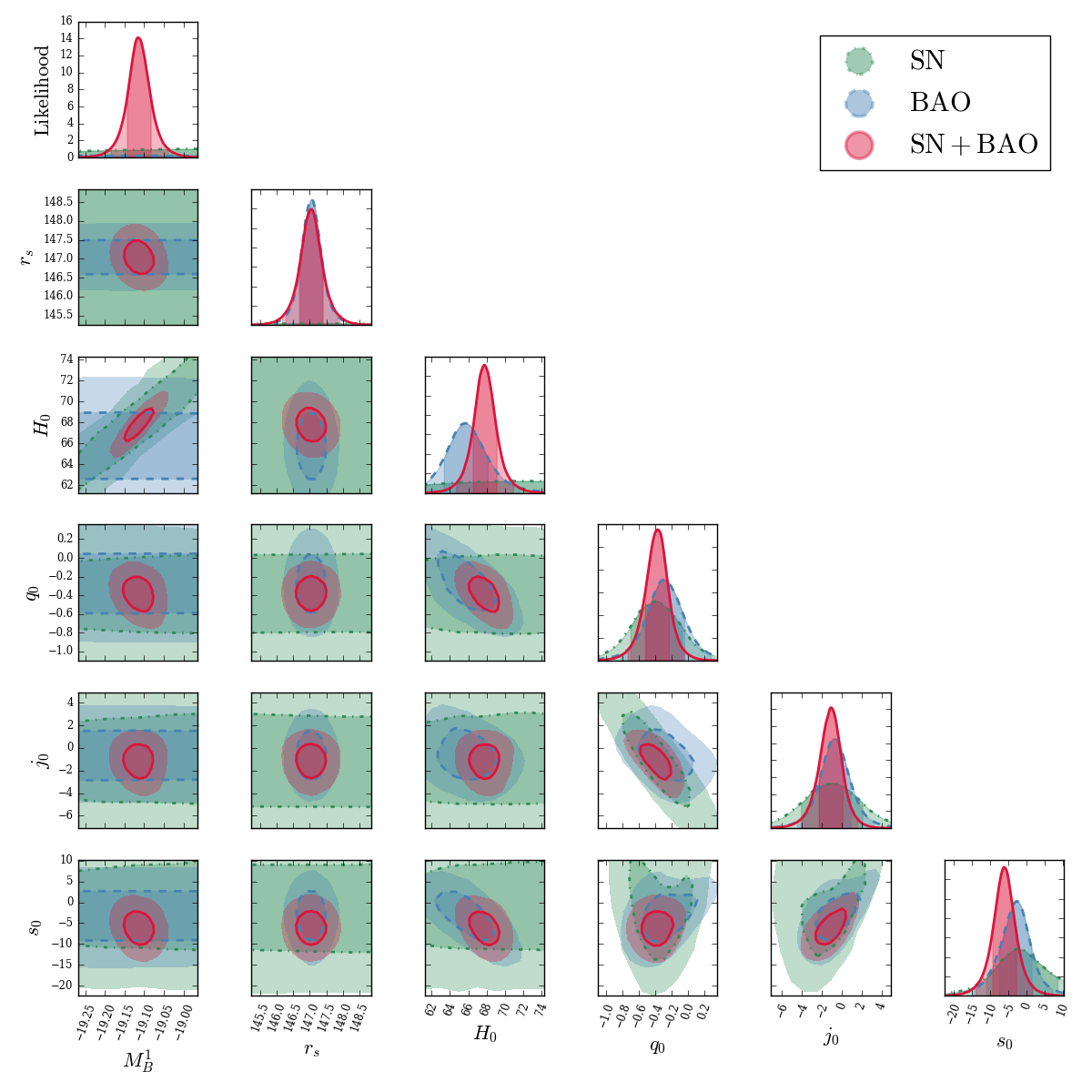}
\caption{The parameter constraints on our model from supernovae (green, dot-dashed lines), BAO (blue, dashed lines), and both data sets combined (red, solid lines).  We can see that the supernovae constraints on $H_0$ and $M^1_B$ are degenerate, and $r_s$ is entirely unconstrained, although the cosmographic parameters of $q_0$, $j_0$ and $s_0$ (which affect the shape of the Hubble diagram) are well constrained.  We can also see that the BAO-only constraint on $H_0$ is correlated with all of the cosmographic parameters.  The power of the combined fit is driven by the degree of orthogonality of the individual constraints, particularly in the $H_0$-$M^1_B$ plane.}. 
   \label{fig:triangle_plot}
\end{center}
\end{figure*}

We use \texttt{emcee} \citep{emcee} as our Markov chain Monte Carlo sampler to determine the joint likelihood of our parameters ($H_0$, $q_0$, $j_0$, $s_0$, $r_s$, $M^1_B$) in Equation \ref{eq:likelihood_combined}. These joint likelihoods are shown in Figure \ref{fig:triangle_plot} along with the marginalised likelihood functions for all the fitted parameters. 

We blinded our analysis throughout the analysis to reduce confirmation bias \citep[e.g.,][]{2011arXiv1112.3108C}. This blinding has been achieved by preparing and testing all our codes and plots using either simulated DES-SN3YR samples (see Appendix \ref{appendix:reproduce} for details) or the existing JLA sample from \cite{2014A&A...568A..22B}. We only replaced these testing data files with the genuine DES-SN3YR sample before submission to the DES collaboration for internal collaboration review. Before unblinding, we reviewed the results with unknown random offsets added to the chains, so that the shape of the likelihoods and the uncertainties could be assessed without influence from the maximum likelihood values of the chains.  After unblinding, some minor updates to the SN-data covariance matrix were introduced, including the use of two different intrinsic dispersion values for the DES and Low-$z$ samples.  Updating our results after unblinding did not significantly change our results or conclusions.

In Table \ref{tab:results} we summarise our $H_0$ measurements with different supernovae surveys.  We note that the $\chi^2 / $ DoF increases for all of the combined fits with SNe Ia data.  This is consistent with the higher $H_0$ values in the combined fits than the BAO-only case, and suggests that the best-fit cosmographic model favours increased expansion at lower redshifts than the BAO surveys.

We note that our cosmographical method can reproduce the $H_0$ measurement from \cite{2017MNRAS.470.2617A}.  Using the same data-set (namely, JLA supernovae, BOSS DR12 BAOs, and the older BAO measurements from \cite{2011MNRAS.416.3017B} and \cite{2015MNRAS.449..835R}), we find $H_0=67.1\pm 1.3$ km s$^{-1}$ Mpc$^{-1}$.  We do not use the BAO measurements from \cite{2011MNRAS.416.3017B} and \cite{2015MNRAS.449..835R} elsewhere, since they are superseded by the combined analysis of \cite{carter18}.
 
We find a larger uncertainty than \cite{2017MNRAS.470.2617A}, which may be due to the more general model we are using for the redshift-distance relation.  Instead of the cosmographic model used here, \cite{2017MNRAS.470.2617A} use a `PolyCDM' cosmological model, which is the $\Lambda$CDM model, plus two additional arbitrary density components, $\Omega_1$ and $\Omega_2$, which scale with linear and quadratic order with redshift, in order to allow for deviations from $\Lambda$CDM.

We find that our best-fit parameters are consistent with previous analyses. For example, we find \textcolor{black}{$M_B^1=-19.12\pm0.03$} mag (68\% confidence) which is consistent with the value from \cite{2015PhRvD..92l3516A} of $M_B^1=-19.14\pm0.042$ mag. Moreover, Figure \ref{fig:triangle_plot} illustrates that, as expected, SNe Ia alone are unable to constrain this parameter, given the strong correlation with $H_0$ (also seen in the figure).  

The cosmographical parameters are constrained by both BAO and SNe Ia. The deceleration parameter shows a significant, negative value of $q_0=-0.37\pm0.15$ (68\% confidence), even after marginalising over other parameters. This value is consistent with other $q_0$ measurements in the literature (e.g., \cite{Lamp2010} found $q_0 = -0.34\pm0.18$ from just the SDSS SN sample) but less negative than expected for a $\Lambda$--dominated cosmology \citep[e.g., $q_0=-0.55$ from ][]{2008PhRvD..78f3504C}. Other cosmographical parameters ($j_0$, $s_0$) are best constrained by the SNe Ia data (green contours), and consistent with zero and expectations from $\Lambda$CDM given our uncertainties. 

We find $H_0= \Ho$ km s$^{-1}$ Mpc$^{-1}$ (68\% confidence), with a reduced $\chi^2$ of 1.16.  This measurement is consistent with the Planck value of $H_0= 67.4 \pm 0.5$ km s$^{-1}$ Mpc$^{-1}$, but inconsistent at 2.5 $\sigma$ to the recent \cite{2018ApJ...855..136R} local distance ladder measurement. 

We make no further comment on this tension as there is already significant literature \citep[e.g., see ][]{freeman17} on the possible systematic uncertainties involved in all measurements and/or interesting new physics that could be responsible. However, we emphasise the independence of this result to that of \cite{2018ApJ...855..136R}: although both measurements rely on SNe Ia, the Cepheid-calibrated distance ladder method of \cite{2018ApJ...855..136R} is very different to the BAO-calibrated inverse distance ladder method used here.  Our measurement is in excellent agreement with previous inverse distance ladder measurements, e.g., \cite{2017MNRAS.470.2617A} who used the JLA sample, as well more recent measurements using the Pantheon SN sample \citep{Feeney18,lemos18}.

\begin{center}
\begin{table}

\begin{tabular}{l | c c c c }
 & $H_0$ &   &  &  \\
Data & [ km s$^{-1}$ Mpc$^{-1}$] & DoF  & $\frac{\chi^2}{\rm{DoF}}$ & $p$-value \\
\hline
BAO Only & 65.7 $\pm$ 2.4 & 2 & 0.75 & 0.47  \\
+JLA & 67.1 $\pm$ 1.3 & 33 & 1.28 & 0.13 \\
+Low-$z$ & 66.1 $\pm$ 2.6 & 10 & 0.94 & 0.49 \\
+DES & 66.9 $\pm$ 1.9 & 12 & 1.28 &  0.22 \\
+DES+Low-$z$ & $\Ho$ & 20  & 1.16 & 0.28 \\
\end{tabular}
\caption{A summary of our $H_0$ measurements.  We fit all the data sets with 6 free parameters (except for the BAO-only case, with 5 free parameters, since we do not fit for $M_B$). }  
\label{tab:results}
\end{table}
\end{center}

\begin{figure*}
\begin{center}
 \includegraphics[width=18cm]{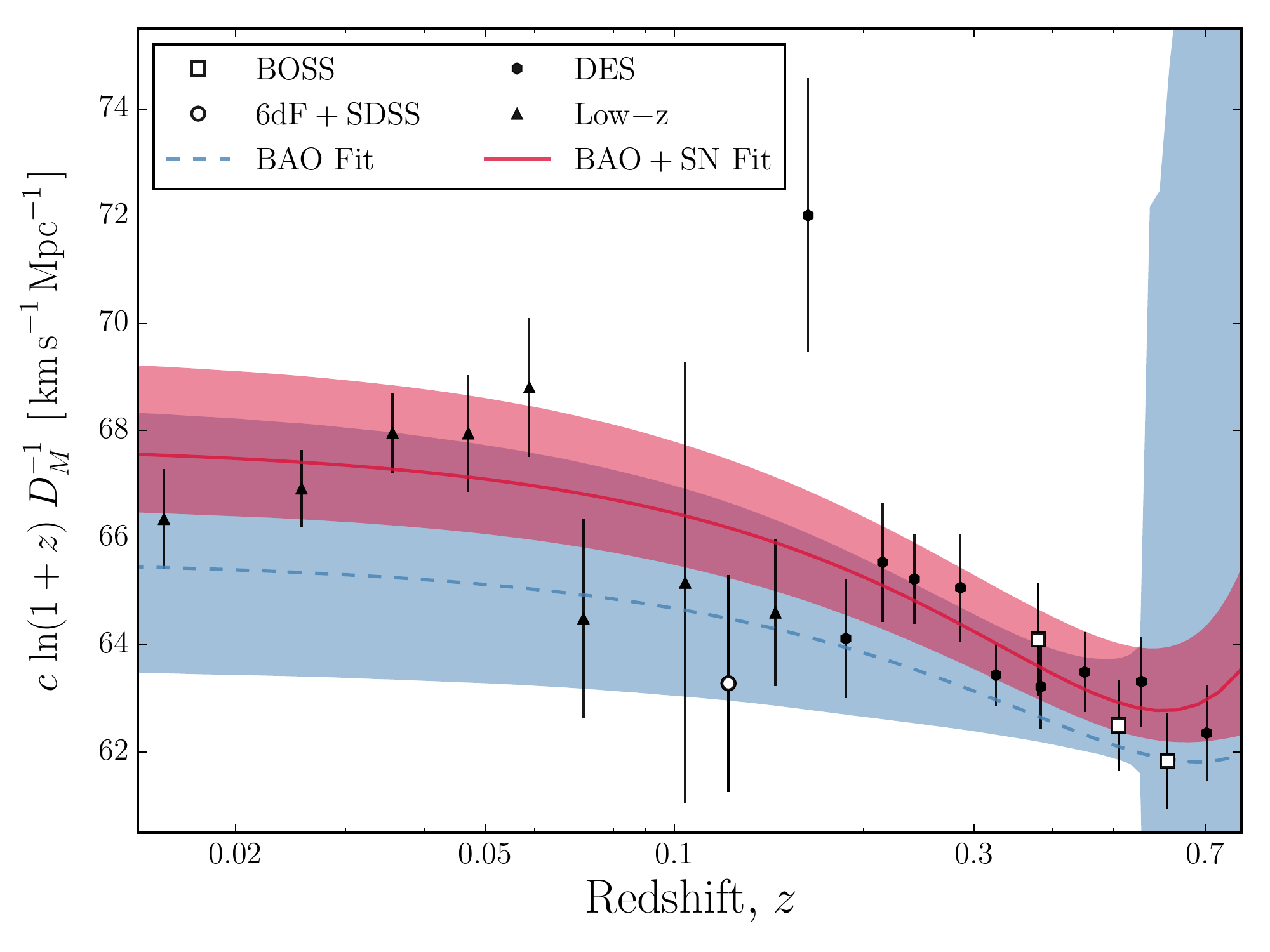}
\caption{Here we illustrate the inverse distance ladder method.  The white data points are the BAO distance measurements, and the black data points are the SNe Ia data.  The DES-SN3YR sample comprises the higher redshift DES SNe (illustrated with hexagonal points), and the Low-$z$ SNe (illustrated with triangular points).  The red line shows our best-fit cosmographical model, and the shaded region is the 68\%  confidence region. The blue dashed line and shaded region illustrates the equivalent constraints from just the BAO data, without any supernovae.  The blue, BAO-only region is very large at $z>0.7$ because we fit only for $r_s$, and not the absolute distance scale at the CMB.}
   \label{fig:inv_dist_ladder}
\end{center}
\end{figure*}

In Figure \ref{fig:inv_dist_ladder}, we provide an illustration of the inverse distance ladder method and the importance of both BAO and SNe Ia data for deriving the best constraint on $H_0$. The SNe Ia data is the DES-SN3YR sample (18 redshift bins) and the SN and BAO uncertainties are the square roots of the corresponding diagonal elements of their covariance matrices. We also show our best-fit cosmographical model (with associated 68\% confidence band in red) as well as the best-fit BAO-only cosmographical model and uncertainty band (in blue).

\section{Discussion \& Conclusions}
\label{sec:Discussion}

Although our $H_0$ value is in excellent agreement with \cite{2018arXiv180706209P}, we emphasise that the use of an $r_s$ prior from Planck does not imply that our measured value of $H_0$ will inevitably agree with the value of $H_0$ derived from Planck cosmological parameters assuming a $\Lambda$CDM cosmology.  

The value of $r_s$ is informed by only the baryon and matter densities at $z=1090$; there are many viable cosmological models which are consistent with only these two quantities (or, in other words, this value of $r_s$) that have wildly different values of the Hubble constant at $z=0$.

Indeed, our BAO-only value of $H_0= 65.7 \pm 2.4$ km s$^{-1}$ Mpc$^{-1}$ is lower than the Planck-derived value.  The BAO data only directly constrain the expansion down to $z\sim0.12$. Extrapolating to $z=0$ thus leads to larger uncertainties from the BAO-only than from the BAO+SNe Ia because the SNe Ia probe down to $z~0.015$.  The consistency between our measurement and the derived Planck value is instead a reflection of the consistency between the cosmology traced by the SN and BAO data and the model used to derive the Planck value.

\subsection{Systematic Uncertainty Budget}
\label{sec:sys_budget}

We now consider the possible systematic uncertainties that may affect our result.  As in \cite{2018arXiv181102377B}, we consider contributions to the total uncertainty from many contributions of systematic uncertainty.  To quantify the effect of each systematic uncertainty, we first repeat our analysis with only the statistical uncertainties included in the supernova data covariance matrix, to find the statistical-only uncertainty on $H_0$,  $\sigma_{\rm{stat}}$.  We then repeat this analysis for each of the sources of systematic uncertainty, including each contribution of systematic uncertainty in the supernova covariance matrix, to find the combined uncertainty due to statistics and the particular systematic,  $\sigma_{\rm{stat+sys}}$.  We then define the systematic only uncertainty as 
\begin{equation}
\sigma_{\rm{sys}} = \sqrt{   \sigma^2_{\rm{stat+sys}}     -\sigma^2_{\rm{stat}}        }.
\label{eq:sys_Error}
\end{equation}

We consider systematic uncertainties due to DES and Low-$z$ calibration, SALT fitting, Supercal calibration \citep{2015ApJ...815..117S}, the intrinsic scatter model, colour parameter parent populations, volume limits, peculiar velocities, flux uncertainty, spectroscopic efficiency, the use of reference cosmology in validation simulations, the Low-$z$ sample 3$\sigma$ outlier cut, the parent population uncertainty, PS1 Coherent Shift, and the use of two different values of the intrinsic dispersion for the DES and Low-$z$ samples \citep{2018arXiv181102377B}.  For each of these systematics, in Table \ref{tab:sytbudget} we quote the shift in the $H_0$ value, and also the fractional contribution of the systematic compared to the statistical uncertainty.  

We find that the total systematic uncertainty from each of these contributions is 72\% of the statistical uncertainty in our measurement; variations below the statistical error are still significant when comparing analysis of the same data.  We note that the various individual systematic uncertainties will not necessarily add in quadrature to the total systematic uncertainty, since each systematic introduces a different weighting of the redshift bins.

\begin{center}
\begin{table}
\begin{tabular}{l | c c c }
Description & $H_0$ shift & $\sigma_{\rm{syst}}$ & $\sigma_{\rm{syst}}$ / $\sigma_{\rm{stat}}$ \\
\hline
Total Stat. & \, 0.000 & 1.048 & 1.00 \\
Total Sys. & 0.162 & 0.760 & 0.72 \\
\hline 
\\
 ALL Calibration & -0.078 & 0.375 & 0.36 \\
 \, DES Cal. & -0.016 & 0.276 & 0.26 \\
 \, Low-$z$ Cal & -0.026 & 0.254 & 0.24 \\
 \, SALT & 0.053 & 0.217 & 0.21 \\
 \\
 ALL Other & 0.004 & 0.661 & 0.63 \\
 \, Intrinsic Scatter & 0.129 & 0.330 & 0.31 \\
 \, $z+0.00004$ & 0.036 & 0.083 & 0.08 \\
 \, $c,x_1$ Parent Pop. & -0.031 & 0.249 & 0.24 \\
 \, Low-$z$ Vol. Lim. & -0.081 & 0.124 & 0.12 \\
 \, Flux Err. & -0.004 & 0.179 & 0.17 \\
 \, Spec. Eff & -0.091 & 0.125 & 0.12 \\
 \, Ref. Cosmo. & -0.065 & 0.134 & 0.13 \\
\, Low-$z$ 3$\sigma$ Cut & 0.498 & 0.193 & 0.18 \\
\, Sys. Parent & 0.370 & 0.222 & 0.21 \\
\, PS1 Coherent Shift & 0.064 & 0.246 & 0.23 \\
\, 2 $\sigma_{\rm{int}}$ & -0.068 & 0.231 & 0.22 \\
\end{tabular}
\caption{The contributions of systematic uncertainties on our $H_0$ measurement. For each term we quantify the shift in the value of $H_0$, the value of the systematic uncertainty, and the fraction of the uncertainty compared to the statistical uncertainty.  The individual uncertainties will not necessarily sum in quadrature to the total uncertainty, due to different redshift weightings and correlations in each term.}
\label{tab:sytbudget}
\end{table}
\end{center}

\begin{figure*}
\begin{center}
 \includegraphics[width=18cm]{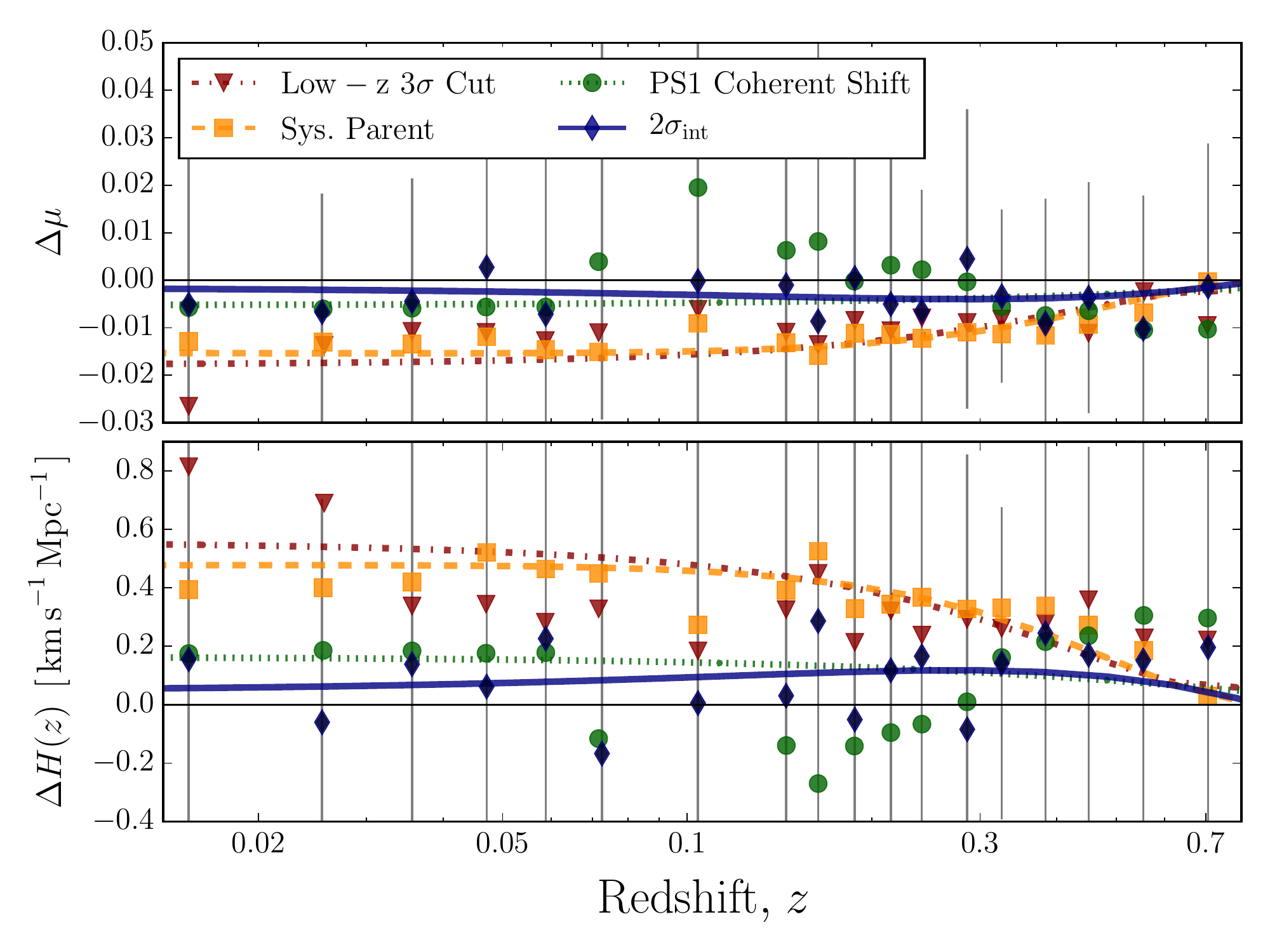} 
\caption{Illustrating the effect of systematic uncertainties.  In the upper panel we plot the change in distance modulus, $\Delta \mu$.  The supernova data are shown with points, and the best-fit models are shown with lines.  For each systematic, we plot the residual compared to the best-fit model and data with statistical only uncertainties.  In the lower panel, we plot the inferred $H(z)$ residual, again subtracting the statistical-only result.  The light-grey vertical lines are the 68\% confidence error bars on the data (which we have plotted for only the PS1 Coherent Shift systematic, for clarity).  The systematics are described in Section \ref{sec:sys_budget}.}
   \label{fig:systs_plot}
\end{center}
\end{figure*}

In Figure \ref{fig:systs_plot}, we illustrate the effect of the Low-$z$ sample 3$\sigma$ outlier cut, the parent population uncertainty, a shift in the absolute wavelength calibration of DES (PS1 Coherent Shift), and the 2$\sigma_{\rm{int}}$ systematic uncertainties.  In the upper panel, we illustrate the effect on the observed distances, $\mu$, relative to the statistical-only distances. In the lower panel, we illustrate the corresponding shift in $H_0$ (again, relative to the statistical-only result).

\subsection{BAO Calibration}
\label{ap:BAO}

One uncertainty in our analysis is our assumed prior on the sound horizon, based on the Planck CMB measurement. The measurement of $r_s$ is very similar between Planck 2013, 2015 and 2018, and changing between the measurements has a negligible effect on our results. Moreover, \cite{2018ApJ...853..119A} and \cite{lemos18} showed that the $H_0$ tension was still present using non-Planck CMB data for $r_s$ (e.g., WMAP+SPT) or measurements of the primordial deuterium abundance from damped Lyman-$\alpha$ systems to constrain the baryon density, and thus sound horizon, independent of any measurement of the CMB power spectrum.

That said, the existence of `dark radiation', or an additional relativistic species in the pre-recombination era, could affect the sound horizon. The most obvious candidate would be massive neutrinos, but constraints on such particles are becoming increasingly tight from a combination of the CMB with measurements of the large-scale structure. For example, \cite{2017JCAP...06..047Y} provide a constraint on the sum of the neutrino masses of $\sum m_{\mu} < 0.8$ eV (95\% confidence) with just data from the Lyman-$\alpha$ forest (LyAF). This constraint improves to  $\sum m_{\mu} < 0.14$ eV (95\% confidence) when combined with the CMB power spectrum, although adding in the recent DES Y1 clustering analysis \citep{rozo17} relaxes these constraints to $\sum m_{\mu} < 0.29$ eV.

\begin{figure*}
\begin{center}
 \includegraphics[width=18cm]{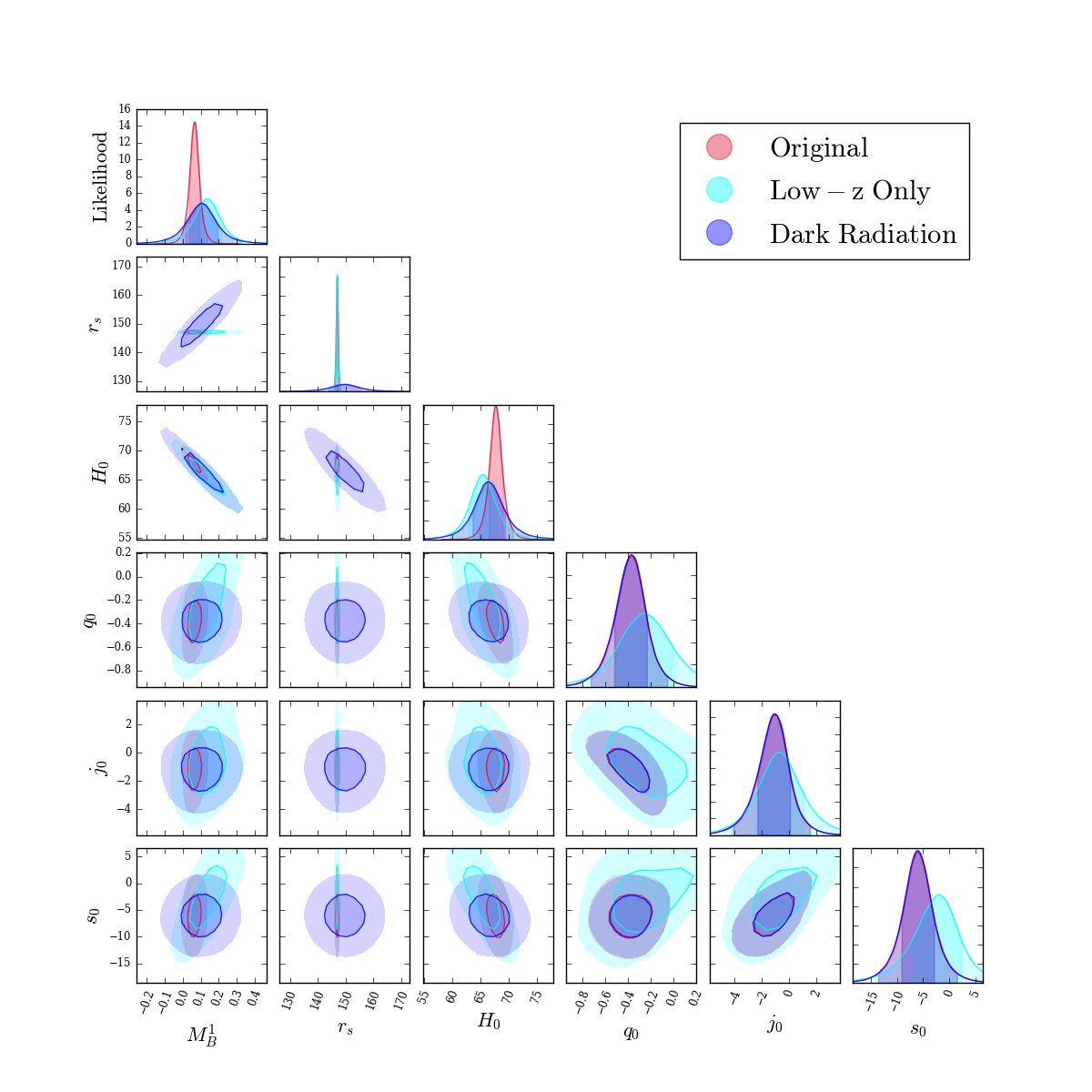}
\caption{Testing the sensitivity of our parameters to different priors and data cuts.  The red contours show our original combined constraints from Figure \ref{fig:triangle_plot}.  In cyan we show the constraints using just the Low-$z$ sample, and in blue we show the effect of varying our prior on the sound horizon, to allow for the possibility of early dark radiation.}
   \label{fig:sys_test_plot}
\end{center}
\end{figure*}

\cite{2017JCAP...04..023V} calculated the effect on $r_s$ due to all remaining observationally allowed contributions of dark radiation, and found $r_s = 150 \pm 5$ Mpc. We therefore test our sensitivity to the possibility of early dark radiation by repeating our analysis with this wider prior on $r_s$. Our results are shown in Figure \ref{fig:sys_test_plot}, and we find $H_0=66.32 \pm 2.9$ km s$^{-1}$Mpc$^{-1}$ (compared to $H_0= \Ho$ km s$^{-1}$ Mpc$^{-1}$ based on our original prior on $r_s$). As expected, this wider prior increases the uncertainty on $H_0$, making it more consistent with the value from \cite{2018ApJ...855..136R}.  We do not include this uncertainty in our uncertainty budget, since the evidence for early dark radiation is not well established, but provide the value for comparison.

As a final test of the sound horizon, we remove any prior on the sound horizon and fit for $r_s$ as a free parameter.  Even without any prior on the value of $r_s$, we are able to place some constraints on $r_s$ (and, correspondingly, $H_0$) with the minimal assumption that $r_s$ is the same for each of the BAO measurements.  

While this assumption would be insufficient in the case of a single (volume averaged) BAO measurement, having multiple BAO measurements at different redshift ranges will -- in principle -- determine $r_s$ (modulo any uncertainty in the cosmographic parameters).  Moreover, in the case of 2D BAO measurements, the consistency of $r_s$ in the parallel and perpendicular measurements further constrains $r_s$ with the Alcock-Paczynski effect.  In other words, the value of $r_s$ is -- in principle -- over-determined, up-to the uncertainties in the cosmographic parameters (which are themselves constrained by the BAO measurements, and also independently by the SN data).

With no prior on $r_s$, we find $r_s=145.2\pm18.5$ Mpc, which is close to the Planck value, although with a much greater uncertainty, which reduces our sensitivity to $H_0$, leading to a value of $H_0=68.9 \pm 8.9$ km s$^{-1}$Mpc$^{-1}$. This test illustrates the importance of knowing the absolute scale of the sound horizon, while also indicating that future BAO measurements from galaxy redshift surveys (e.g., Euclid and DESI) should help constrain this parameter independent of the CMB and thus remove any reliance on early universe plasma physics.

One possible concern is the assumption of a fiducial cosmology in converting the galaxy angular positions and redshifts observed in a galaxy redshift survey (such as BOSS) into the power spectrum of galaxy clustering where the BAO signal is determined. As stated in Section \ref{sec:BAO}, this issue is addressed by assuming a scaling law (Equation \ref{eq:Dv}) which `dilates' the distance being tested to the fiducial cosmology used to calculate the galaxy power spectrum. The applicability of such scaling was first studied in detail by \cite{PW2008} who showed, using $N$-body simulations, a systematic uncertainty of only $\simeq1\%$ on $\alpha^{\rm{BAO}}$ over a wide range of $\alpha^{\rm{BAO}}$ values, or more importantly, over a wide range of alternative cosmologies.  

It is also worth stressing that the BAO signal is estimated in a series of narrow redshift bins, thus allowing for uncertainties between the assumed and fiducial cosmology to be minimised across any single redshift shell (assuming cosmologies with a smooth redshift-distance relationship close to $\Lambda$CDM). The effects of such redshift binning has recently been examined by \cite{2016MNRAS.461.2867Z} using mock galaxy catalogues that closely mimic BOSS, and they found their BAO analysis and measurements remained unbiased even when the assumed fiducial cosmology differed from the true (simulation) cosmology. They also confirmed an uncertainty of just $1$\% on $\alpha^{\rm{BAO}}$ over a range of different assumptions (fiducial cosmologies, pivot redshifts, redshift-space distortion parameters, and galaxy bias models) which could be further improved to the sub-percent level with future optimal redshift weighting schemes \citep[e.g.,][]{2015MNRAS.451..236Z}. For reference, \cite{2017MNRAS.470.2617A} assumed a 0.3\% systematic uncertainty on $\alpha^{\rm{BAO}}$ from their fitting methodologies.

We note that the redshifts at which the BAO measurements are made are approximate, because they are weighted averages of the redshifts of all of the pairs of galaxies that go into generating the correlation function.  The weighting can depend on the choice of using average redshifts or average distances, which adds some uncertainty to the redshift of the distance anchor.  However, since the slope in Figure \ref{fig:inv_dist_ladder} is shallow, any uncertainty in the centre of the redshift bin would only add a small uncertainty to the measurement of $H_0$ ($<$ 0.5 km s$^{-1}$Mpc$^{-1}$).

Therefore, the systematic uncertainties on the BAO measurements should be subdominant at present, and should not affect our conclusions given the larger statistical uncertainties on our $H_0$ results.

\subsection{Comparison to \protect\cite{2015PhRvD..92l3516A} and \protect\cite{2017MNRAS.470.2617A} }

In this section, we compare our results to earlier inverse distance ladder measurements by \cite{2015PhRvD..92l3516A} and \cite{2017MNRAS.470.2617A}, using the JLA supernova sample.  \cite{2015PhRvD..92l3516A} found $H_0= 67.3 \pm 1.1$ km s$^{-1}$ Mpc$^{-1}$ with BAOs from BOSS DR11, which changed only marginally to  $H_0= 67.3 \pm 1.0$ km s$^{-1}$ Mpc$^{-1}$ in  \cite{2017MNRAS.470.2617A} with BOSS DR12.  We note that these values are consistent with our value of  $H_0= \Ho$ Mpc$^{-1}$ using the DES-SN3YR sample, and  $H_0= 67.1 \pm 1.3$ km s$^{-1}$ Mpc$^{-1}$ using JLA.

We emphasize a number of differences in the methods to these papers.  \cite{2015PhRvD..92l3516A} and \cite{2017MNRAS.470.2617A} assumed a model for the redshift-distance relation based on $\Lambda$CDM, with the addition of two additional, arbitrary density components, which scaled to linear and quadratic order with the scale factor, $a$, in order to allow for the possibility of physics beyond $\Lambda$CDM.  In order to constrain the BAO scale, $r_s$, these papers applied priors on matter and baryon densities derived from the CMB, and used a fitting function to calculate $r_s$ as a function of these parameters \citep[see][]{2015PhRvD..92l3516A}.

In contrast, our method is more physics-agnostic.  The cosmographic model we assume means that we do not have to assume a Friedmann model for the redshift-distance relation.  The prior on the scale of $r_s$ also means that we are less sensitive to the (albeit, well understood) plasma physics of the CMB.  

As for data, we additionally use the BAO measurement from \cite{carter18}.  As illustrated in Figure \ref{fig:inv_dist_ladder}, we believe that the inclusion of this BAO measurement leads to our marginally lower value of $H_0$ with the JLA sample: 67.1, vs. 67.3 km s$^{-1}$ Mpc$^{-1}$ in   \cite{2015PhRvD..92l3516A} and \cite{2017MNRAS.470.2617A}.

Although our value of $H_0$ has a larger uncertainty than in \cite{2017MNRAS.470.2617A} (1.3 vs. 1.0  km s$^{-1}$ Mpc$^{-1}$), this is due to our more physics-agnostic model, as opposed to the data.  Comparing both sets of SNe directly with our cosmographic model, we find the same uncertainty with DES-SN3YR as with JLA (1.3 km s$^{-1}$ Mpc$^{-1}$), even though DES-SN3YR comprises 329 SNe, compared to 740 in JLA. 

Moreover, we also include several additional sources of systematic uncertainty which were not included by \cite{2014A&A...568A..22B}.  For example, we allow for a systematic uncertainty due to a redshift-shift caused by large-scale inhomogeneities \citep{2017JCAP...01..038C}.  We also note in Figure \ref{fig:sys_test_plot} that the uncertainty in the underlying parent population of SNe Ia is one of the dominant sources of systematic uncertainty \citep[e.g.,][]{2010A&A...523A...7G,2011ApJS..192....1C}.

This smaller uncertainty on $H_0$ is driven in part by the BBC analysis method \citep{2017ApJ...836...56K}, and also by the calibration consistency of DECam in the DES SN sample \citep{2015AJ....150..150F,2018arXiv181102377B}.  We note, however, that since DES-SN3YR and JLA share some SNe in the Low-$z$ sample, they cannot be combined, as the samples are not entirely independent.

\section{Acknowledgements}

We thank the anonymous referee for comments and suggestions which have considerably improved this paper.  We thank An\u{z}e Slosar for help and advice with the method and likelihood code, and Paul Carter and Florian Beutler for help with the BAO data. 

E.M., R.C.N., D.B. acknowledge funding from STFC grant ST/N000668/1. TC acknowledges the University of Portsmouth for a Dennis Sciama Fellowship. D.S. is supported by NASA through Hubble Fellowship grant HST-HF2-51383.001 awarded by the Space Telescope Science Institute, which is operated by the Association of Universities for Research in Astronomy, Inc., for NASA, under contract NAS 5- 26555. We acknowledge funding from ERC Grant 615929.

A.V.F. is grateful for financial assistance from NSF grant AST-1211916, the Christopher R. Redlich Fund, the TABASGO Foundation, and the Miller  Institute for Basic Research in Science (U.C. Berkeley).

The UCSC team is supported in part by NASA grants 14-WPS14-0048, NNG16PJ34G, NNG17PX03C, NSF grants AST-1518052 and AST-1815935, the Gordon \& Betty Moore Foundation, the Heising-Simons Foundation, and by fellowships from the Alfred P.\ Sloan Foundation and the David and Lucile Packard Foundation to R.J.F.

This paper makes use of observations taken using the Anglo-Australian Telescope under programs ATAC A/2013B/12 and  NOAO 2013B-0317; the Gemini Observatory under programs NOAO 2013A-0373/GS-2013B-Q-45, NOAO 2015B-0197/GS-2015B-Q-7, and GS-2015B-Q-8; the Gran Telescopio Canarias under programs GTC77-13B, GTC70-14B, and GTC101-15B; the Keck Observatory under programs U063-2013B, U021-2014B, U048-2015B, U038-2016A; the Magellan Observatory under programs CN2015B-89; the MMT under 2014c-SAO-4, 2015a-SAO-12, 2015c-SAO-21; the South African Large Telescope under programs 2013-1-RSA\_OTH-023, 2013-2-RSA\_OTH-018, 2014-1-RSA\_OTH-016, 2014-2-SCI-070, 2015-1-SCI-063, and 2015-2-SCI-061; and the Very Large Telescope under programs ESO 093.A-0749(A), 094.A-0310(B), 095.A-0316(A), 096.A-0536(A), 095.D-0797(A).

Funding for the DES Projects has been provided by the U.S. Department of Energy, the U.S. National Science Foundation, the Ministry of Science and Education of Spain, 
the Science and Technology Facilities Council of the United Kingdom, the Higher Education Funding Council for England, the National Center for Supercomputing 
Applications at the University of Illinois at Urbana-Champaign, the Kavli Institute of Cosmological Physics at the University of Chicago, 
the Center for Cosmology and Astro-Particle Physics at the Ohio State University,
the Mitchell Institute for Fundamental Physics and Astronomy at Texas A\&M University, Financiadora de Estudos e Projetos, 
Funda{\c c}{\~a}o Carlos Chagas Filho de Amparo {\`a} Pesquisa do Estado do Rio de Janeiro, Conselho Nacional de Desenvolvimento Cient{\'i}fico e Tecnol{\'o}gico and 
the Minist{\'e}rio da Ci{\^e}ncia, Tecnologia e Inova{\c c}{\~a}o, the Deutsche Forschungsgemeinschaft and the Collaborating Institutions in the Dark Energy Survey. 

The Collaborating Institutions are Argonne National Laboratory, the University of California at Santa Cruz, the University of Cambridge, Centro de Investigaciones Energ{\'e}ticas, 
Medioambientales y Tecnol{\'o}gicas-Madrid, the University of Chicago, University College London, the DES-Brazil Consortium, the University of Edinburgh, 
the Eidgen{\"o}ssische Technische Hochschule (ETH) Z{\"u}rich, 
Fermi National Accelerator Laboratory, the University of Illinois at Urbana-Champaign, the Institut de Ci{\`e}ncies de l'Espai (IEEC/CSIC), 
the Institut de F{\'i}sica d'Altes Energies, Lawrence Berkeley National Laboratory, the Ludwig-Maximilians Universit{\"a}t M{\"u}nchen and the associated Excellence Cluster Universe, 
the University of Michigan, the National Optical Astronomy Observatory, the University of Nottingham, The Ohio State University, the University of Pennsylvania, the University of Portsmouth, 
SLAC National Accelerator Laboratory, Stanford University, the University of Sussex, Texas A\&M University, and the OzDES Membership Consortium.

Based in part on observations at Cerro Tololo Inter-American Observatory, National Optical Astronomy Observatory, which is operated by the Association of Universities for Research in Astronomy (AURA) under a cooperative agreement with the National Science Foundation.

The DES data management system is supported by the National Science Foundation under Grant Numbers AST-1138766 and AST-1536171.
The DES participants from Spanish institutions are partially supported by MINECO under grants AYA2015-71825, ESP2015-66861, FPA2015-68048, SEV-2016-0588, SEV-2016-0597, and MDM-2015-0509, 
some of which include ERDF funds from the European Union. IFAE is partially funded by the CERCA program of the Generalitat de Catalunya.
Research leading to these results has received funding from the European Research
Council under the European Union's Seventh Framework Program (FP7/2007-2013) including ERC grant agreements 240672, 291329, and 306478.
We  acknowledge support from the Australian Research Council Centre of Excellence for All-sky Astrophysics (CAASTRO), through project number CE110001020, and the Brazilian Instituto Nacional de Ci\^encia
e Tecnologia (INCT) e-Universe (CNPq grant 465376/2014-2).

This manuscript has been authored by Fermi Research Alliance, LLC under Contract No. DE-AC02-07CH11359 with the U.S. Department of Energy, Office of Science, Office of High Energy Physics. The United States Government retains and the publisher, by accepting the article for publication, acknowledges that the United States Government retains a non-exclusive, paid-up, irrevocable, world-wide license to publish or reproduce the published form of this manuscript, or allow others to do so, for United States Government purposes.

\bibliographystyle{mnras}
\bibliography{references}

\appendix

\section{Tests of the Parameter-Fitting Code}
\label{appendix:reproduce}

In this section, we test our method with artificially generated SN and BAO distances, to ensure we can recover the input parameters used to generate the distances.  To generate the artificial distances, we use our cosmographic model with values chosen at $H_0 = 70.0$ km s$^{-1}$ Mpc$^{-1}$, $q_0 = -0.55$, $j_0 = -0.85$, $s_0 = -0.9$, $M^1_B = -19.07$, and $r_s = 147$ Mpc.    

To generate the artificial supernova distances, we calculated fiducial distance moduli $\mu^{\rm{fid}}(z)$ at the redshift values of the genuine supernova data.  To generate artificial distance moduli with a realistic dispersion, we drew realisations from a correlated Gaussian distribution centred on $\mu_{\rm{fid}}$ with covariance given by the genuine data covariance matrix.  

We took a similar approach to generating artificial distances for the BOSS DR12 BAO measurements for this cosmographic model.  We first calculated a vector of fiducial $H^{\rm{fid}}(z)$ and $D^{\rm{fid}}_m(z)$ at the three effective redshift bins of the BAO measurements. To generate realisations of the BAO data, we then drew samples from a correlated Gaussian distribution centred on this observation vector with covariance of the genuine DR12 data covariance matrix.  

\begin{figure*}
\begin{center}
 \includegraphics[width=18cm]{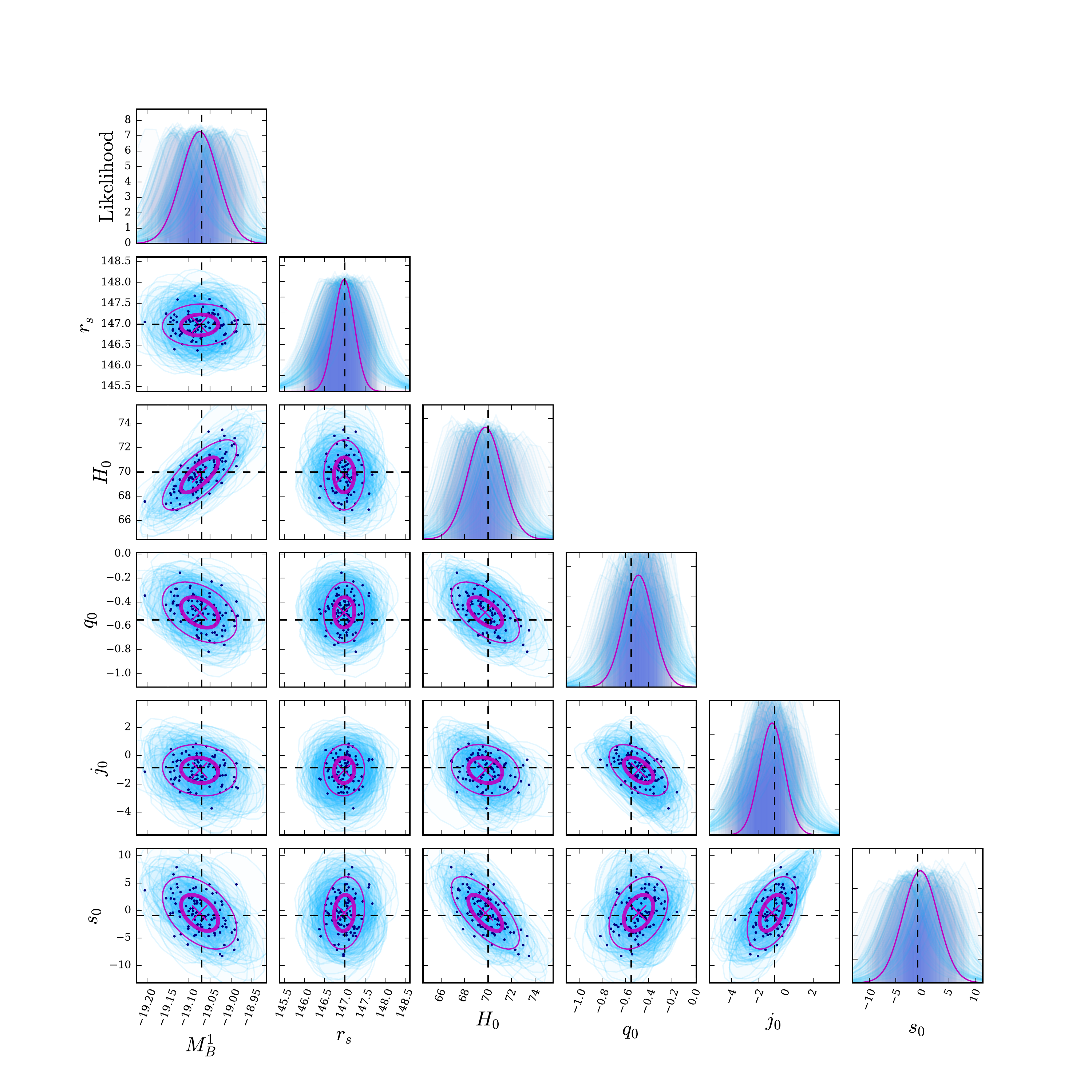}
\caption{The best-fit parameters for 100 mock realisations (see text for details). The black dashed lines show the input parameter values for our mock realisations, while the light blue contours show the 68\% confidence region for each of the hundred realisations.  The dark blue points show the maximum likelihood values for each realisation. The magenta ellipses are the one and two standard deviations of these best-fit points, centred on the magenta crosses, at the averages of the individual maximum likelihood values.}
   \label{fig:sim_test_plot}
\end{center}
\end{figure*}

In total, we generated one hundred artificial realisations of DES-SN3YR and BOSS DR12 data-sets. These simulations are shown in Figure \ref{fig:sim_test_plot} along with the average of these realisations (contours).  These simulations confirm we can accurately recover the input parameters, and the uncertainties on the simulated parameter measurements are consistent with our genuine uncertainties.

\section{Author Affiliations}
\label{appendix:affiliations}

$^{1}$ Institute of Cosmology and Gravitation, University of Portsmouth, Portsmouth, PO1 3FX, UK\\
$^{2}$ Department of Physics and Astronomy, University of Pennsylvania, Philadelphia, PA 19104, USA\\
$^{3}$ School of Mathematics and Physics, University of Queensland,  Brisbane, QLD 4072, Australia\\
$^{4}$ ARC Centre of Excellence for All-sky Astrophysics (CAASTRO)\\
$^{5}$ The Research School of Astronomy and Astrophysics, Australian National University, ACT 2601, Australia\\
$^{6}$ African Institute for Mathematical Sciences, 6 Melrose Road, Muizenberg, 7945, South Africa\\
$^{7}$ South African Astronomical Observatory, P.O.Box 9, Observatory 7935, South Africa\\
$^{8}$ Kavli Institute for Cosmological Physics, University of Chicago, Chicago, IL 60637, USA\\
$^{9}$ Department of Astronomy and Astrophysics, University of Chicago, Chicago, IL 60637, USA\\
$^{10}$ Lawrence Berkeley National Laboratory, 1 Cyclotron Road, Berkeley, CA 94720, USA\\
$^{11}$ School of Physics and Astronomy, University of Southampton,  Southampton, SO17 1BJ, UK\\
$^{12}$ Cerro Tololo Inter-American Observatory, National Optical Astronomy Observatory, Casilla 603, La Serena, Chile\\
$^{13}$ Fermi National Accelerator Laboratory, P. O. Box 500, Batavia, IL 60510, USA\\
$^{14}$ Korea Astronomy and Space Science Institute, Yuseong-gu, Daejeon, 305-348, Korea\\
$^{15}$ LSST, 933 North Cherry Avenue, Tucson, AZ 85721, USA\\
$^{16}$ Department of Physics \& Astronomy, University College London, Gower Street, London, WC1E 6BT, UK\\
$^{17}$ George P. and Cynthia Woods Mitchell Institute for Fundamental Physics and Astronomy, and Department of Physics and Astronomy, Texas A\&M University, College Station, TX 77843,  USA\\
$^{18}$ Kavli Institute for Particle Astrophysics \& Cosmology, P. O. Box 2450, Stanford University, Stanford, CA 94305, USA\\
$^{19}$ SLAC National Accelerator Laboratory, Menlo Park, CA 94025, USA\\
$^{20}$ Centro de Investigaciones Energ\'eticas, Medioambientales y Tecnol\'ogicas (CIEMAT), Madrid, Spain\\
$^{21}$ Laborat\'orio Interinstitucional de e-Astronomia - LIneA, Rua Gal. Jos\'e Cristino 77, Rio de Janeiro, RJ - 20921-400, Brazil\\
$^{22}$ INAF, Astrophysical Observatory of Turin, I-10025 Pino Torinese, Italy\\
$^{23}$ Department of Astronomy, University of Illinois at Urbana-Champaign, 1002 W. Green Street, Urbana, IL 61801, USA\\
$^{24}$ National Center for Supercomputing Applications, 1205 West Clark St., Urbana, IL 61801, USA\\
$^{25}$ Institut de F\'{\i}sica d'Altes Energies (IFAE), The Barcelona Institute of Science and Technology, Campus UAB, 08193 Bellaterra (Barcelona) Spain\\
$^{26}$ Institut d'Estudis Espacials de Catalunya (IEEC), 08034 Barcelona, Spain\\
$^{27}$ Institute of Space Sciences (ICE, CSIC),  Campus UAB, Carrer de Can Magrans, s/n,  08193 Barcelona, Spain\\
$^{28}$ Observat\'orio Nacional, Rua Gal. Jos\'e Cristino 77, Rio de Janeiro, RJ - 20921-400, Brazil\\
$^{29}$ Department of Astronomy/Steward Observatory, 933 North Cherry Avenue, Tucson, AZ 85721-0065, USA\\
$^{30}$ Jet Propulsion Laboratory, California Institute of Technology, 4800 Oak Grove Dr., Pasadena, CA 91109, USA\\
$^{31}$ Department of Astronomy, University of Michigan, Ann Arbor, MI 48109, USA\\
$^{32}$ Department of Physics, University of Michigan, Ann Arbor, MI 48109, USA\\
$^{33}$ Department of Astronomy, University of California, Berkeley, CA 94720-3411, USA\\
$^{34}$ Miller Senior Fellow, Miller Institute for Basic Research in Science, University of California, Berkeley, CA  94720, USA\\
$^{35}$ Santa Cruz Institute for Particle Physics, Santa Cruz, CA 95064, USA\\
$^{36}$ PITT PACC, Department of Physics and Astronomy, University of Pittsburgh, Pittsburgh, PA 15260, USA\\
$^{37}$ Instituto de Fisica Teorica UAM/CSIC, Universidad Autonoma de Madrid, 28049 Madrid, Spain\\
$^{38}$ Centre for Astrophysics \& Supercomputing, Swinburne University of Technology, Victoria 3122, Australia\\
$^{39}$ CENTRA, Instituto Superior T\'ecnico, Universidade de Lisboa, Av. Rovisco Pais 1, 1049-001 Lisboa, Portugal\\
$^{40}$ Department of Physics, ETH Zurich, Wolfgang-Pauli-Strasse 16, CH-8093 Zurich, Switzerland\\
$^{41}$ Center for Cosmology and Astro-Particle Physics, The Ohio State University, Columbus, OH 43210, USA\\
$^{42}$ Department of Physics, The Ohio State University, Columbus, OH 43210, USA\\
$^{43}$ Max Planck Institute for Extraterrestrial Physics, Giessenbachstrasse, 85748 Garching, Germany\\
$^{44}$ Universit\"ats-Sternwarte, Fakult\"at f\"ur Physik, Ludwig-Maximilians Universit\"at M\"unchen, Scheinerstr. 1, 81679 M\"unchen, Germany\\
$^{45}$ Harvard-Smithsonian Center for Astrophysics, Cambridge, MA 02138, USA\\
$^{46}$ Department of Physics, University of Namibia, 340 Mandume Ndemufayo Avenue, Pionierspark, Windhoek, Namibia\\
$^{47}$ Australian Astronomical Optics, Macquarie University, North Ryde, NSW 2113, Australia\\
$^{48}$ Sydney Institute for Astronomy, School of Physics, A28, The University of Sydney, NSW 2006, Australia\\
$^{49}$ Departamento de F\'isica Matem\'atica, Instituto de F\'isica, Universidade de S\~ao Paulo, CP 66318, S\~ao Paulo, SP, 05314-970, Brazil\\
$^{50}$ Department of Astronomy, The Ohio State University, Columbus, OH 43210, USA\\
$^{51}$ Instituci\'o Catalana de Recerca i Estudis Avan\c{c}ats, E-08010 Barcelona, Spain\\
$^{52}$ Division of Theoretical Astronomy, National Astronomical Observatory of Japan, 2-21-1 Osawa, Mitaka, Tokyo 181-8588, Japan\\
$^{53}$ Institute of Astronomy and Astrophysics, Academia Sinica, Taipei 10617, Taiwan\\
$^{54}$ Department of Physics and Astronomy, Pevensey Building, University of Sussex, Brighton, BN1 9QH, UK\\
$^{55}$ Brandeis University, Physics Department, 415 South Street, Waltham MA 02453\\
$^{56}$ Instituto de F\'isica Gleb Wataghin, Universidade Estadual de Campinas, 13083-859, Campinas, SP, Brazil\\
$^{57}$ Computer Science and Mathematics Division, Oak Ridge National Laboratory, Oak Ridge, TN 37831\\
$^{58}$ Observatories of the Carnegie Institution for Science, 813 Santa Barbara St., Pasadena, CA 91101, USA\\
$^{59}$ Argonne National Laboratory, 9700 South Cass Avenue, Lemont, IL 60439, USA\\

\bsp	
\label{lastpage}
\end{document}